\documentstyle[prl,aps,floats]{revtex}
\input{epsf}
\input{psfig}

\font\latin=cmsy10
\def\guz{\hbox{\latin P$_{\rm G}$}}

\def\be{\begin{equation}}
\def\ee{\end{equation}}
\def\bea{\begin{eqnarray}}
\def\eea{\end{eqnarray}}

\def\ea{{\sl et	al., }}
\def\ean{{\sl et al.\ }}
\def\tj{$t$--$J$\ }

\begin{document}
\draft
\twocolumn[\hsize\textwidth\columnwidth\hsize\csname@twocolumnfalse%
\endcsname

\title{
The two-leg $t$--$J$ ladder: a spin liquid
generated by Gutzwiller	projection of magnetic bands}

\author{Martin Greiter}
\address{Department of Physics,	Stanford University, Stanford, CA 94305,
greiter@quantum.stanford.edu}

\date{SU-ITP 98/19, cond-mat/9804002, March 31, 1998}

\maketitle

\begin{abstract}
The ground state of the	two-leg	Heisenberg ladder is identified	as an
RVB type spin liquid, which is generated by Gutzwiller projection of
tight-binding bands with flux $\pi$ per	plaquet.
Explicit trial wave functions for the magnon and hole excitations
are formulated in terms	of spinons and holons.
\end{abstract}
]

1.\
Itinerant antiferromagnets confined to coupled chains, or $t$--$J$ ladders,
have enjoyed enormous popularity over the past few years
\cite{science,troyer,poil,white,shelton,sofive,soeight,nagaosa,gagli}.
They provide the simplest example of a generic spin	liquid in dimensions
greater	than one, and the only example thereof which is	presently
fully amenable to numerical methods.  (It is furthermore widely	believed
that they constitute the first step towards understanding the two-dimensional
$t$--$J$ model starting	from one dimension, but	I rather believe
the two-leg ladder to be just a	special	case.)
These models are approximately realized	in
(VO)$_2$P$_2$O$_7$, SrCu$_2$O$_3$, and Sr$_{14-x}$Ca$_x$Cu$_{24}$O$_{41}$,
and hence accessible to	experiment.

The \tj	Hamiltonian for	the ladder is given by
\be
H_{t-J}
= - \sum_{\langle ij\rangle\, \sigma} t_{ij} c^\dagger_{i\sigma}c_{j\sigma}
+{1\over 2}\sum_{\langle ij\rangle} J_{ij}\, {\bf S}_i\cdot{\bf	S}_j
\label{eq:tjh}
\ee
where $(t_{ij},J_{ij})=(t,J)$ if $i$ and $j$ are nearest neighbors
along one of the chains, and $(t_\perp,J_\perp)$ if they are nearest neighbors
across the rungs; each pair $\langle ij\rangle$
is summed over twice and no doubly occupied sites are allowed.

One of the most	striking features of the two-leg $t$--$J$ ladder is
the persistence	of a spin gap $\Delta \approx J_\perp/2$
in the weak coupling limit $J_\perp \ll	J$.
(For sufficiently strong couplings $J_\perp > J$,
the system can be described by a perturbative expansion
around the strong coupling limit
consisting of singlets across the rungs\cite{gap}, which yields	a spin gap
$\Delta\approx J_\perp-J+{1\over2}J^2/J_\perp $;
a weak coupling	expansion starting from	decoupled chains, however,
is not possible, as the	individual spin	chains are quantum critical
in the sense that the tiniest perturbation can
change the universality	class.)
In this	letter,	I will formulate a microscopic theory of the two-leg
\tj ladder, which is universally valid at {\it all ratios} $J_\perp/J$,
in terms of explicit spin liquid trial wave functions for the ground state,
magnon ({\it spinon-spinon} bound state) and the hole ({\it holon-spinon}
bound state) excitations.

\begin{figure}
\epsfbox{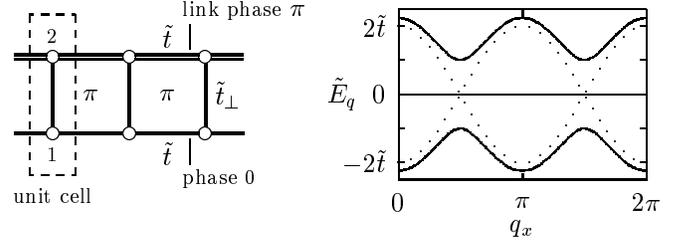}
\vspace{10pt}
\caption{Flux band structure of	a tight	binding	ladder with
flux $\pi$ per plaquet for
$\tilde	t_\perp/\tilde t=0$ (doted lines) and $\tilde t_\perp/\tilde t=1$
(solid lines).	The energy gap is given	by $2\tilde t_\perp$.
}
\label{fig:three}
\end{figure}

2.\
The trial wave function	for the	ground state of	the Heisenberg
ladder,	the \tj	ladder without any holes, is constructed as follows.
Consider a tight binding ladder	with flux $\pi$	per
plaquet, and hopping terms of magnitude	$\tilde	t$ along the chains, and
$\tilde	t_\perp$ across	the rungs.  In the gauge depicted in
figure \ref{fig:three},	we write the single particle Bloch states
\be
\psi_{{	q}}(j) = e^{i {{ q}} \cdot {{ R}}_j}
u_{{ q}}(j),
\ee
where the $u_{{	q}}(j)$	are strictly periodic
in both	real and momentum space
and obey
\be
\tilde H_{{ q}}	\left( \begin{array}{c}
			    u_{{ q}}(1)	\\ u_{{	q}}(2)
			 \end{array} \right) =
\tilde E_{{ q}}	\left( \begin{array}{c}
			    u_{{ q}}(1)	\\ u_{{	q}}(2)
			 \end{array} \right)
\label{eq:dirac}
\ee
where
\be
\tilde H_{{ q}}	= 2\tilde t \left( \begin{array}{cc}
				 \cos q_x & m e^{-iq_y}	\\
				 m e^{iq_y} & -\cos q_x
			      \end{array} \right)
\ee
with $m\equiv\tilde t_\perp/2\tilde t$.	 Since (\ref{eq:dirac})	is a
two-dimensional	Dirac equation,	i.e.\ $\tilde H_{{ q}}^2$ is
diagonal, we immediately obtain	the eigenvalues
\be
\tilde E_{{ q}}=\pm 2\tilde t \sqrt{\cos^2 q_x + m^2}.
\ee
The coupling between the tight-binding chains hence induces an
energy gap of magnitude	$2\tilde t_\perp$.
We now fill the	lower band twice, once with up-spin electrons,
and once with down-spin	electrons; the resulting 
Slater determinant $|\psi_{\rm SD}\rangle$
is obviously a spin singlet.  The spin liquid trial wave function
for the	Heisenberg ladder with
$J_\perp/J=\tilde t_\perp/\tilde t$ is obtained	by
eliminating all	the doubly occupied sites via Gutzwiller projection,
\be
|\psi_{\rm trial}\rangle = \guz	|\psi_{\rm SD}\rangle.
\ee
Since the Gutzwiller projector \guz\ commutes with the total spin operator,
$|\psi_{\rm trial}\rangle$ is also a singlet.  This trial wave function	is
as accurate an approximation as	the Haldane-Shastry state\cite{hs} for the
one-dimensional	Heisenberg chain in the	weak coupling limit $J_\perp/J=0$,
and exact in the strong	coupling limit $J_\perp/J\rightarrow\infty$;
the approximation has its worst	point at isotropic coupling
(see table \ref{tab:four}) \cite{finite,opt}.

\begin{table}
\begin{tabular}{c|rrccrrrr}
$J_\perp/J$ &
\multicolumn{2}{c}{$E_{\rm tot}$} &\% &over- &
\multicolumn{2}{c}{${\langle \vec S_i \vec S_j \rangle}_\parallel $} &
\multicolumn{2}{c}{${\langle \vec S_i \vec S_j \rangle}_\perp $} \\
& exact	& trial	& off &	lap & exact & trial & exact & trial \\ \hline
 0   & -9.031 &	-9.015 & 0.2 & 0.997 &-0.452 &-0.451 &0.000  &0.000 \\
 0.1 & -9.062 &	-9.024 & 0.4 & 0.986 &-0.450 &-0.451 &-0.062 &-0.011\\
 0.2 & -9.155 &	-9.073 & 0.9 & 0.969 &-0.445 &-0.449 &-0.123 &-0.045\\
 0.5 & -9.755 &	-9.568 & 1.9 & 0.952 &-0.420 &-0.413 &-0.269 &-0.263\\
 1 &-11.577 &-11.346 & 2.0 & 0.941 &-0.354 &-0.302 &-0.450 &-0.530\\
 2 & -8.594 & -8.444 & 1.8 & 0.957 &-0.222 &-0.143 &-0.638 &-0.702\\
 5 & -7.664 & -7.594 & 0.9 & 0.981 &-0.085 &-0.029 &-0.732 &-0.748\\
10 & -7.539 & -7.513 & 0.3 & 0.993 &-0.040 &-0.007 &-0.746 &-0.750\\
$\infty$ & -7.500 & -7.500 & 0.0 & 1.000 &0.000	&0.000 &-0.750 &-0.750
\end{tabular}
\vspace{5pt}
\caption{Energy	expectation values and nearest neighbor
spin correlations for the spin liquid trial wave functions in comparison
with the exact ground states of	a $2\times 10$ Heisenberg
ladder with periodic boundary conditions, as well as
overlaps between trial wave functions and exact	ground states.
Throughout this	article, all energies quoted are in units of
max($J_\perp$,$J$).  The boundary phase	
before Gutzwiller projection has been $0$.}
\label{tab:four}
\end{table}

3.\
There are essentially two ways of constructing spinon and holon
excitations for	spin liquids (they are obtained	from each other	by
annihilating or	creating an electron on	the spinon or holon site).
The first one is Anderson's projection
technique\cite{phil}: inhomogenities in	both spin
and charge created before Gutzwiller projection	yield inhomogenities
in spin	only after projection.	Anderson writes	a state	with two
spinons	localized at sites $i$ and $j$
\be
|\psi_{i\uparrow ,j\downarrow} \rangle =
\guz c^\dagger_{i\uparrow}c_{j\uparrow}|\psi_{\rm SD}\rangle.
\ee
In the case of the ladder, however, the	spinons	are not	free particles,
but bound into pairs by	a linear confinement force
\cite{ffc}.  To	obtain the magnon trial	wave function,
\be
|\psi_{\rm magnon}({k})\rangle = \sum_{i,j} \phi_{i,j}({k})
|\psi_{i\uparrow ,j\uparrow} \rangle,
\ee
a hence	nontrivial internal wave function $\phi_{i,j}({k})$ for	the
spinon-spinon bound state is required.	(In
the one-dimensional spin chain,	by contrast, the spinons interact
only weakly, and the internal spinon wave function can be approximated
by plane waves.)

It is therefore	expedient to use the second method, which has been
successful in describing the fractionally charged solitons in
polyacytelene\cite{poly}:
Rokhsar\cite{rokh}\ constructs
elementary excitations of spin liquids via localized midgap states,
which are either occupied by
a single electron (spinon) or left unoccupied (holon).	The topology
of the ladder dictates that midgap states can only be created in pairs,
which implies that we automatically obtain spinon-spinon (or holon)
bound states rather then isolated spinons (or holons).
The fact that the energy required to create two	singly occupied
midgap states is proportional to $\tilde t_\perp$ suggests a spin gap
proportional to	$J_\perp$.
These general observations, however, leave us still with
a large	number of possible choices for the midgap states; most
constructions yield satisfactory magnon, but only very
few acceptable hole trial wave functions.
To identify those, let us step back and	take a broader view.

\begin{figure}
\epsfbox{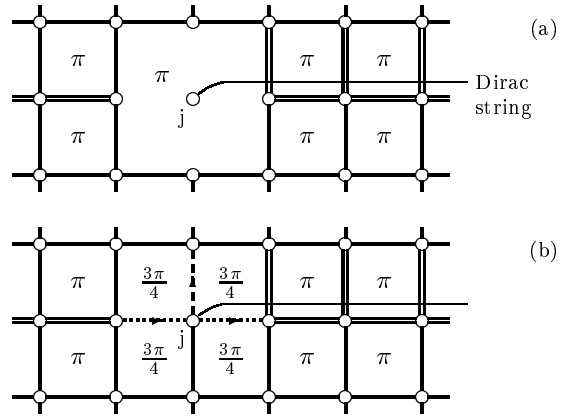}
\vspace{10pt}
\caption{Magnetic tight-binding	configuration for a holon of the
chiral spin liquid (a) as proposed by Rokhsar
(cutting all the links to a given site and adjusting the flux according
to Rokhsar's loop rules	generates a stationary holon or	holetto) or
(b) by combining this flux adjustment with Anderson's projection technique
(only the phases of the	hopping	parameters around the holon site are
adjusted).
}
\label{fig:ten}
\end{figure}

4.\
The spin liquid	proposed
above is, in fact, a special case of the
Kalmeyer-Laughlin chiral spin liquid\cite{chiral}, obtained
by imposing a periodic boundary condition with a periodicity of only two
lattice	spacings in $y$-direction. 
This chiral spin liquid	may be generated from a	tight-binding lattice
with flux $\pi$	per plaquet and
hopping	magnitudes $\tilde t$ and ${1\over2}\tilde t_\perp$
in $x$-	and $y$-direction, respectively;
the P and T violating diagonal hopping elements,
which are otherwise required to	open an	energy gap, 
cancel due to the boundary condition.

Spinons	and holons for the chiral spin liquid may be constructed via
Anderson's method or via midgap	states;	Rokhsar	
creates	a midgap state in the flux band	structure before projection
by cutting all the links
to a given site	and adjusting the flux according to his	loop rules,
which require that the kinetic energy on the loops around
each plaquet is	minimal	(see figure \ref{fig:ten}a).  The
resulting holon	
is not nearly as mobile	as Anderson's, but optimal with	regard to the
magnetic energy; it adequately describes stationary charge excitations.
I call it a stationary holon or	{\it holetto}.
To obtain the generic and mobile holon,	we create a midgap state by adjusting
the flux according to Rokhsar's	procedure
(i.e.\ we create a defect of flux $\pi$	around the holon site)
without	cutting	any links
(i.e.\ we adjust the hopping phases without adjusting the magnitudes),
and then project such that this	site
is unoccupied (see figure \ref{fig:ten}b).  This holon is equivalent to
Anderson's
in the case of the chiral spin liquid, but more generally applicable.

\begin{figure}
\epsfbox{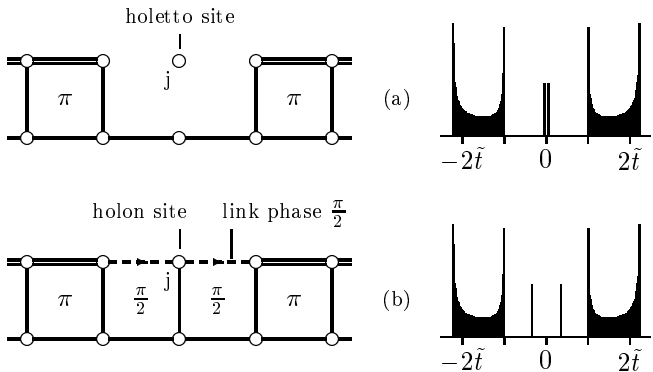}
\vspace{10pt}
\caption{Magnetic tight-binding	configurations and the corresponding
density	of states for the ladder with a	spinon bound to	(a)
a stationary {\it holetto} or (b) a mobile {\it	holon}
before Gutzwiller projection.  Only the	latter
flux configuration violates P and T.}
\label{fig:seven}
\end{figure}

5.\
The flux configurations	used to	construct
holetto-spinon and holon-spinon	bound states
for the	ladder
are shown in figure \ref{fig:seven}a and b.  In	the case of the	holetto,
we cut all the links to	a given	site; as the topology of the ladder
does not provide a context for a flux adjustment around	this site,
we obtain a second midgap state, and hence a spinon, localized nearby.
This trial wave	function describes a stationary	hole.
To construct a mobile hole,
we create the midgap states by only adjusting the flux,	and
project	such that the holon site is unoccupied;	as we are creating two
rather than one	midgap state, we remove	flux $\pi/2$
from each neighboring plaquet\cite{flux}.  The flux configuration now
violates P and T, and the holon-spinon bound state carries
a chirality quantum number, which is $+$ for the configuration
shown in figure	\ref{fig:seven}b, and $-$ for its complex conjugate;
states of opposite chiralities map into	each other under P or T.
The final trial	wave functions for the hole is
a linear superposition of the holon-spinon bound states
of both	chiralities at each momentum,
\be
|\psi_{\rm hole}({k})\rangle = \hbox{\latin N}
\sum_{j} e^{i {	r}_j { k}}
\left(|\psi^+_j	\rangle	+ a({k}) |\psi^-_j \rangle \right),
\ee
where ${r}_j$ is the holon cordinate and $a({k})$ is a variational
parameter.  Magnons or spinon-spinon bound states are obtained from the
holon-spinon bound states by creating an electron at the holon site,
which forms a spin triplet with	the spinon bound to it.

\begin{figure}
\epsfbox{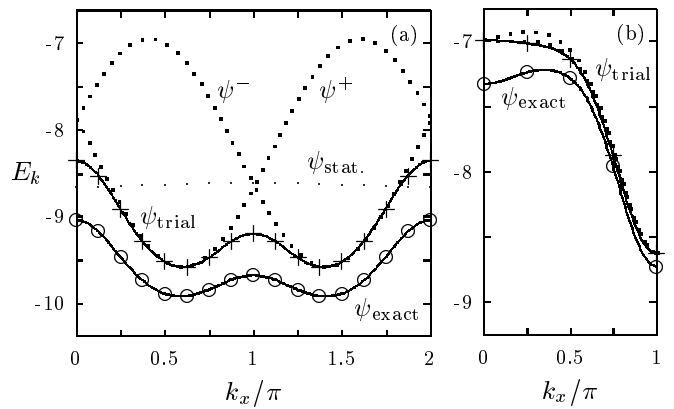}
\vspace{10pt}
\caption{Dispersions for a single (a) hole ({\it holon-spinon} bound state)
and (b)	magnon ({\it spinon-spinon} bound state)
as predicted by	the spin liquid	proposed here in comparison with the
exact dispersions for a	$2\times 8$ ladder with	$J_\perp=J=t_\perp=t=1$
(see tables \ref{tab:one} and \ref{tab:three})
and periodic (or antiperiodic) boundary	conditions.
The dotted lines correspond to
the individual $+$ and $-$ chirality trial
wave functions (generated from the tight-binding configuration shown
in figure \ref{fig:seven}b and its P or	T conjugate, respectively) and
the stationary holetto-spinon bound state (according to	figure
\ref{fig:seven}a), as indicated.
}
\label{fig:eight}
\end{figure}

Numerical comparisons of the trial wave	functions for holes and
magnons	with the exact eigenstates are presented in
figure \ref{fig:eight} and in tables
\ref{tab:one}--\ref{tab:two}.
The hole trial wave functions for 
$t_\perp/J_\perp=t/J=1$	(see table \ref{tab:one})
are excellent at stong coupling, and
less accurate at isotropic coupling; for weak coupling,	they are excellent
only at	momenta	close to the one-hole ground states, 
as there is a large amplitude to find a	hole and a magnon
rather than just a hole	at other momenta.
The holon-spinon bound state wave functions adequately describe	the hole
when $t$ and $J$ are comparable
(see table \ref{tab:two}); for $t\ll J$, the holetto-spinon bound
state is more appropriate, while holes with sufficiently large $t$
are detrimental	to antiferromagnetic correlations and
eventually destabilize the spin	liquid\cite{nagaoka}.
The trial wave functions for the magnons are generally 
satisfactory (see table	\ref{tab:three}).

\begin{table}
\begin{tabular}{cc|rrccrrrrc}
\multicolumn{2}{l|}{$J_\perp/J$} &
\multicolumn{2}{c}{$E_{\rm tot}$} &\% &over- &
\multicolumn{2}{c}{$E_{t_\parallel}$} &
\multicolumn{2}{c}{$E_{t_\perp}$} &
$|a|$ \\
\multicolumn{2}{r|}{\hspace{5mm}$k_x/\pi$} & exact & trial &
\hspace{.7mm}off\hspace{.7mm} &	lap & exact & trial & exact & trial & \\ \hline
    &$0$     & -7.40& -6.88& 7.1& 0.560& -1.51&	-0.45&	0.04&  0.06& 1.00\\
    &${1/4}$ & -7.92& -7.83& 1.1& 0.975& -1.63&	-1.62&	0.01&  0.04& 0.07\\
 0.2&${1/2}$ & -8.17& -8.07& 1.2& 0.952& -1.89&	-1.87& -0.02&  0.04& 0.10\\
    &${3/4}$ & -7.59& -7.22& 4.9& 0.540& -1.49&	-1.01& -0.10&  0.04& 0.08\\
    &$1$     & -7.21& -5.98&  17& 0.000& -1.55&	 0.45& -0.09&  0.06& 1.00\\ \hline
    &$0$     & -9.04& -8.36& 7.6& 0.776& -0.88&	 0.42& -0.38& -0.89& 1.00\\
    &${1/4}$ & -9.46& -8.91& 5.8& 0.849& -1.21&	-0.47& -0.45& -0.67& 0.12\\
 1  &${1/2}$ & -9.89& -9.52& 3.7& 0.894& -1.54&	-1.23& -0.59& -0.57& 0.02\\
    &${3/4}$ & -9.84& -9.47& 3.8& 0.917& -1.35&	-0.99& -0.81& -0.70& 0.16\\
    &$1$     & -9.67& -9.19& 5.0& 0.917& -1.07&	-0.42& -0.98& -0.89& 1.00\\ \hline
    &$0$     & -6.20& -6.12& 1.3& 0.981&  0.13&	 0.19& -0.96& -1.00& 1.00\\
    &${1/4}$ & -6.25& -6.18& 1.1& 0.984&  0.08&	 0.11& -0.97& -0.99& 0.32\\
 5  &${1/2}$ & -6.38& -6.33& 0.7& 0.987& -0.05&	-0.04& -0.98& -0.99& 0.17\\
    &${3/4}$ & -6.50& -6.46& 0.7& 0.988& -0.17&	-0.16& -0.99& -0.99& 0.34\\
    &$1$     & -6.55& -6.50& 0.7& 0.988& -0.21&	-0.19& -1.00& -1.00& 1.00
\end{tabular}
\vspace{5pt}
\caption{Energy	expectation values including individual
contributions to the kinetic energy from chains	and rungs
and overlaps for the trial wavefunctions describing
holon-spinon bound states in comparison	with the exact
one hole eigenstates
for a periodic $2\times	8$ ladder with
$t_\perp/J_\perp=t/J=1$	for three ratios $J_\perp/J$.
The energies are in units
of max($J_\perp$,$J$); the transverse momentum is always $k_y=0$,
as a $k_y=\pi$ state corresponds to a hole (holon-spinon bound state) plus
a magnon (spinon-spinon	bound state).  The trial wave functions
are given by
$|\psi_{\rm trial}\rangle=\hbox{\latin N}
\left( |\psi_+\rangle +	{\text a} |\psi_-\rangle \right) $,
where $|\psi_+\rangle$
is the
trial wave function constructed with the flux configuration
shown in figure	\ref{fig:seven}b and $|\psi_-\rangle$ its P or T
conjugate.
The boundary phase for the flux	band structure before
projection has been $0$	for the	chain containing the holon, and	$\pi$
for the	other chain.
}
\label{tab:one}
\end{table}

\begin{table}
\begin{tabular}{cc|rrccrrrrc}
\multicolumn{2}{l|}{$J_\perp/J$} &
\multicolumn{2}{c}{$E_{\rm tot}$} &\% &over- &
\multicolumn{2}{c}{$E_{J_\parallel}$} &
\multicolumn{2}{c}{$E_{J_\perp}$} &
$|a|$ \\
\multicolumn{2}{r|}{\hspace{5mm}$k_x/\pi$} & exact & trial &
\hspace{.8mm}off\hspace{.8mm} &	lap & exact & trial & exact & trial & \\ \hline
    &$0$     & -6.31& -5.86& 7.2& 0.844& -6.23&	-5.70& -0.09& -0.16& 1.00\\
    &${1/4}$ & -6.16& -6.13& 0.4& 0.994& -6.08&	-6.08& -0.08& -0.05& 0.01\\
 0.2&${1/2}$ & -5.60& -5.56& 0.7& 0.983& -5.44&	-5.47& -0.16& -0.09& 0.04\\
    &${3/4}$ & -5.96& -5.89& 1.1& 0.961& -5.70&	-5.73& -0.26& -0.16& 0.22\\
    &$1$     & -6.98& -6.94& 0.6& 0.990& -6.73&	-6.76& -0.25& -0.18& 1.00\\ \hline
    &$0$     & -7.33& -6.99& 4.6& 0.885& -5.11&	-3.79& -2.22& -3.20& 1.00\\
    &${1/4}$ & -7.24& -7.01& 3.1& 0.927& -4.78&	-3.74& -2.46& -3.27& 0.43\\
 1  &${1/2}$ & -7.28& -7.13& 2.0& 0.961& -4.11&	-3.44& -3.18& -3.69& 0.25\\
    &${3/4}$ & -7.96& -7.87& 1.1& 0.981& -4.56&	-4.22& -3.39& -3.65& 0.10\\
    &$1$     & -8.73& -8.63& 1.1& 0.982& -5.68&	-5.36& -3.04& -3.27& 1.00\\ \hline
    &$0$     & -4.91& -4.86& 1.0& 0.986& -0.04&	 0.12& -4.88& -4.99& 1.00\\
    &${1/4}$ & -4.96& -4.92& 0.9& 0.988& -0.07&	 0.07& -4.89& -4.98& 1.08\\
 5  &${1/2}$ & -5.09& -5.06& 0.6& 0.991& -0.18&	-0.08& -4.91& -4.97& 1.03\\
    &${3/4}$ & -5.24& -5.20& 0.7& 0.991& -0.34&	-0.23& -4.90& -4.98& 1.03\\
    &$1$     & -5.31& -5.26& 0.9& 0.987& -0.43&	-0.28& -4.88& -4.99& 1.00
\end{tabular}
\vspace{5pt}
\caption{As in table \ref{tab:one}, but	now for	spinon-spinon bound states
(magnons) with $k_y=\pi $
obtained by creation of an electron at the holon site.
For $k_x=0$, spinetto-spinon bound states yield	better trial wave functions
(the energy is only 0.4	\% off
at $J_\perp/J=0.2$, and	2.7 \% off at $J_\perp/J=1$).
}
\label{tab:three}
\end{table}

\begin{table}
\begin{tabular}{c|rrccrrrrc}
$t/J$ &
\multicolumn{2}{c}{$E_{\rm tot}$} &\% &over- &
\multicolumn{2}{c}{$E_{t_\parallel}$} &
\multicolumn{2}{c}{$E_{t_\perp}$} &
$|a|$ \\
& exact	& trial	&
\hspace{.7mm}off\hspace{.7mm} &	lap & exact & trial & exact & trial & \\ \hline
 0   & -8.18 & -7.88 & 3.7 & 0.888 &  0.00 &  0.00 &  0.00 &  0.00 & 1.00\\
     &	     & -8.00 & 2.2 & 0.938 &	   &  0.00 &	   &  0.00 & * \\
 0.2 & -8.37 & -8.14 & 2.8 & 0.927 & -0.17 & -0.14 & -0.10 & -0.16 & 0.34\\
     &	     & -8.12 & 2.9 & 0.923 &	   &  0.00 &	   & -0.13 & * \\
 0.5 & -8.87 & -8.63 & 2.7 & 0.929 & -0.65 & -0.53 & -0.28 & -0.33 & 0.12\\
     &	     & -8.32 & 6.3 & 0.862 &	   &  0.00 &	   & -0.32 & * \\
 1   & -9.89 & -9.52 & 3.7 & 0.894 & -1.54 & -1.23 & -0.59 & -0.57 & 0.02\\
 2   &-12.1 &-11.3 & 6.5 & 0.797 & -3.36 & -2.63 & -1.31 & -1.02 & 0.04\\
 5   &-19.5 &-16.8 &  14 & 0.585 & -8.86 & -6.84 & -3.76 & -2.33 & 0.08	
\end{tabular}
\vspace{5pt}
\caption{As in table \ref{tab:one}, but	now with $J_\perp/J=t_\perp/t=1$ for
different ratios $t/J$.	 The momentum is $(k_x,k_y)=(\pi/2,0)$,
which corresponds
to the ground state of the $2\times 8$ ladder with
periodic boundary conditions.  For $t/J\leq 0.5$, data for
holetto-spinon bound states
are shown as well, marked with~*; 
these adequately describe stationary holes ($t=0$).
}
\label{tab:two}
\end{table}

6.\
The P and T violation of the localized
holon-spinon bound states, or the appearance of	a chirality
quantum	number,	is a physical property of the system;
any real trial wave function for the localized hole would yield	a
dispersion $E_{{ k}} \propto \cos (k_x)$,
and thus be inconsistent with the dispersion obtained
by exact diagonalization (bottom curve in
figure \ref{fig:eight}a).  
The chirality quantum number	is a manifestation
of the fractional statistics\cite{wilc}	of spinons and 
holons\cite{chiral,hald} in dimensions greater than one;
it determines the sign of the
statistical phases acquired as they encircle each other.

I am deeply grateful to	Bob Laughlin and Sudip Chakravarty
for many illuminating discussions.
This work was supported	through	NSF grant No.~DMR-95-21888.
Additional support was provided	by the NSF MERSEC Program through
the Center for Materials Research at Stanford University.


\begin{references}
\bibitem{finite}
For a finite ladder with a periodic boundary
condition, the trial wave function depends slightly on the boundary
phase of the flux ladder before	Gutzwiller projection;
it appears that the best trial wave function is obtained when 
the kinetic energy before projection is minimal.
\bibitem{opt}
The trial wave function	can be 
improved by optimizing
$\tilde	t_\perp/\tilde t$ for each value $J_\perp/J$; at isotropic
coupling $J_\perp/J=1$,
for example, the energy	is only	1.35 \%	off if one assumes
$\tilde	t_\perp/\tilde t=0.84$.
\bibitem{flux}
This number, however, is not crucial; removing flux $\pi/4$ from each
neighboring plaquet yields comparable trial wave functions.
\end{references}
\end{document}